\documentclass[prb,twocolumn,showpacs]{revtex4}
\usepackage{graphics}

\newcommand{\Journal}[4]{#1 \textbf{#2}, #3 (#4)}

\newcommand{\be}{\begin{equation}}
\newcommand{\ee}{\end{equation}}

\begin{document}

\title{Comment on ``Epitaxial Pb(Zr,Ti)O$_3$ thin films with coexisting tetragonal
and rhombohedral phases''}
\author{I. A. Sergienko}
\email[E-mail:]{iserg@uic.rsu.ru} \affiliation{Physics Department,
Rostov State University, Rostov-on-Don, 344090, Russia}

%\pacs{77.80.Bh, 77.84.Dy, 81.30.Dz, 77.65.Bn}
%\keywords{}

\maketitle

In a recent paper, Oh and Jang (OJ)~\cite{jang} reported on (001)
PZT thin film, epitaxially grown on a (001) MgO substrate. It was
shown that for compositions Zr/Ti=70/30 and Zr/Ti=80/20, x-ray
diffraction patterns exhibit both tetragonal and rhombohedral
features. Such compositions correspond to rhombohedral Zr-rich
region of the T-$x$ phase diagram of bulk PZT~\cite{jaffe,noh}.
The authors explained the shift of morphotropic phase boundary
(MPB) to Zr-side by the presence of stress caused by different
ferro-/paraelectric as well as thermal expansion properties of the
film and the substrate.

Here we present some corrections to the theory of this effect
proposed by OJ. They assumed that spontaneous strain $x_1$ of the
film-substrate system, under the condition $\lambda_{fc}(T^p-T)\ll
1$, can be expressed as [Eq.~(4) of Ref.~\onlinecite{jang}]

\begin{equation}
\label{oj} x_1\approx(\lambda_{fc}-\lambda_{s})(T^p-T),
\end{equation}
where $\lambda_{fc}$ and $\lambda_{s}$ are thermal expansion
coefficients of film and substrate, respectively, and $T^p$ -- the
processing (annealing) temperature.

We consider stress ($\sigma_i$) related parts of the Gibbs
function ($\Delta G$) for both PZT film and MgO substrate.
Following OJ, we assume that the film is thin enough so that the
stress can be approximated by a spatially uniform average value.
Additionally, we consider the near-boundary layer of the
substrate, omitting the nonhomogenous terms in this case as well.
 For PZT thin film

\begin{eqnarray}
\label{gibbs} \Delta G&=&\lambda_{fc}(T^p-T)
(\sigma_1+\sigma_2+\sigma_3)\nonumber\\
&&-\frac{1}{2}s_{11}(\sigma_1^2+\sigma_2^2+\sigma_3^2)\nonumber\\
&&-s_{12}(\sigma_1\sigma_2+\sigma_2\sigma_3+\sigma_1\sigma_3)\nonumber\\
&&-\frac{1}{2}s_{44}(\sigma_4^2+\sigma_5^2+\sigma_6^2)\nonumber\\
&&-Q_{11}(\sigma_1P_1^2+\sigma_2P_2^2+\sigma_3P_3^2)
-Q_{12}[\sigma_1(P_2^2+P_3^2)\nonumber\\&&+\sigma_2(P_1^2+P_3^2)+\sigma_3(P_1^2+P_2^2)]\nonumber\\
&&-Q_{44}(\sigma_4P_2P_3+\sigma_5P_1P_2+\sigma_6P_1P_3) ,
\end{eqnarray}
where $P_i$ are ferroelectric order parameter components. The rest
of the notations are standard. We would like to emphasize that the
first term in~(\ref{gibbs}) is important since we are interested in
the effects of the thermal expansion~\cite{landau}.

For the cubic paraelectric substrate (primed symbols):

\begin{eqnarray}
\label{gibbs1} \Delta
G'&=&\lambda_{s}(T^p-T)(\sigma'_1+\sigma'_2+\sigma'_3)\nonumber\\
&&-\frac{1}{2}s'_{11}({\sigma'_1}^2+{\sigma'}_2^2+{\sigma'}_3^2)\nonumber\\
&&-s'_{12}({\sigma'}_1\sigma'_2+\sigma'_2\sigma'_3+\sigma'_1\sigma'_3)\nonumber\\
&&-\frac{1}{2}s'_{44}({\sigma'}_4^2+{\sigma'}_5^2+{\sigma'}_6^2),
\end{eqnarray}

If there is no external stress applied, the spontaneous values are
$\sigma_3=\sigma_4=\sigma_5=\sigma'_3=\sigma'_4=\sigma'_5=0$~\cite{landau}.
In contrast to Eq.~(1) of OJ, $\sigma_6$ and $\sigma'_6$ in
general do not vanish because shear strain in the lower symmetry
structure of the film evolves upon cooling, resulting in the
stress from the substrate.

On the other hand, in the tetragonal phase of PZT
$\sigma_6=\sigma'_6=0$ and $\sigma_1=\sigma_2=H$,
$\sigma'_1=\sigma'_2=H'$.

It follows that spontaneous strain components are

\begin{eqnarray}
\label{strain} x_1&=&-\left(\frac{\partial \Delta G}{\partial
\sigma_1}\right)_{T,P_i}=\lambda_{fc}(T-T^p)+(s_{11}+s_{12})H\nonumber\\
&&+Q_{11}P_1^2+Q_{12}(P_2^2+P_3^2)\nonumber\\
x'_1&=&-\left(\frac{\partial \Delta G'}{\partial
\sigma'_1}\right)_T=\lambda_{s}(T-T^p)+(s'_{11}+s'_{12})H',
\end{eqnarray}
where for tetragonal phase, polarized along [001] $P_1=P_2=0,
P_3\ne0$, for rhombohedral phase $P_1=P_2\ne0, P_3\ne0$, and
$P_1\ne P_3$ if $H$ is finite.

\begin{figure}
\includegraphics{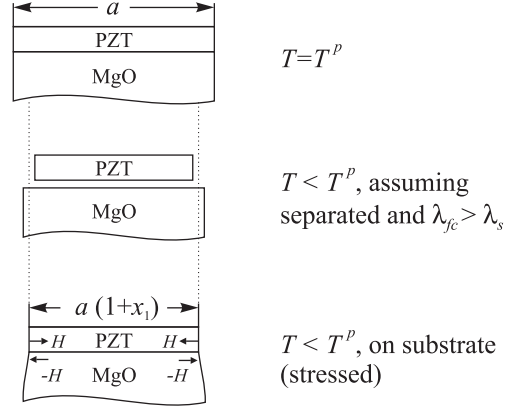}
\caption{\label{fig} Thermal expansion of PZT thin film and MgO
substrate along [100] direction. $T$ -- temperature, $a$ -- film
extent at the processing temperature $T^p$, $x_1$ and $H$ are
strain and stress for $T<T^p$.}
\end{figure}

It is clear from Fig.~\ref{fig}, that at the interface between the film
and the substrate the following boundary conditions are satisfied

\begin{equation}
\label{bound} x_1=x'_1, \qquad H=-H'.
\end{equation}

Now, it is straightforward from Eqs.~(\ref{strain}) and~(\ref{bound})

\begin{eqnarray}
\label{res}
x_1&=&\frac{(s'_{11}+s'_{12})\lambda_{fc}+(s_{11}+s_{12})\lambda_{s}}
{s'_{11}+s'_{12}+s_{11}+s_{12}}(T-T^p)\nonumber\\
&&+\frac{s'_{11}+s'_{12}}{s'_{11}+s'_{12}+s_{11}+s_{12}}
[Q_{11}P_1^2+Q_{12}(P_2^2+P_3^2)],\nonumber\\
H&=&\frac{(\lambda_{fc}-\lambda_s)(T^p-T)}{s'_{11}+s'_{12}+s_{11}+s_{12}}\nonumber\\
&&-\frac{Q_{11}P_1^2+Q_{12}(P_2^2+P_3^2)}{s'_{11}+s'_{12}+s_{11}+s_{12}}.
\end{eqnarray}

As we mentioned above, in rhombohedral phase the spontaneous values
$\sigma_6\ne0$ and $\sigma'_6\ne0$, but they should be defined
self-consistently form Eqs.~(\ref{gibbs}),~(\ref{gibbs1}), and
boundary conditions similar to those in~(\ref{bound})
\begin{eqnarray}
x_6=x'_6&=&\frac{s'_{44}Q_{44}P_1P_3}{s'_{44}+s_{44}}\nonumber\\
\sigma_6=-\sigma'_6&=&-\frac{Q_{44}P_1P_3}{s'_{44}+s_{44}}
\end{eqnarray}

In summary, our approach does not require more than solving linear
equations, if only terms of up to second order in stress are taken
into account in $\Delta G$ and $\Delta G'$. At the same time,
elastic properties of the substrate have to be taken into account
to describe the film-substrate system thermal expansion, and the
corrections to Eq.~(\ref{oj}) appear even in the lowest order
approximation.

The author would like to thank S. Urazhdin for helpful
discussions.

\end{document}